\documentclass[prd,nofootinbib,showpacs,preprint]{revtex4}
\usepackage{amsmath}
\usepackage{graphicx}
\usepackage{epsfig}
\graphicspath{{Figs/}}
\usepackage{dcolumn}
\usepackage{bm}
\usepackage{amssymb}
\usepackage[usenames,dvipsnames]{color}
\usepackage{slashed}
\usepackage[dvipdfm,colorlinks,citecolor=blue]{hyperref}
\begin{document}
\title{Quasi-Degenerate Neutrinos in Type II Seesaw Models}
\author{Mrinal Kumar Das}
\email{mkdas@tezu.ernet.in}
\affiliation{Department of Physics, Tezpur University, Tezpur-784028, India}
\author{Debasish Borah}
\email{dborah@tezu.ernet.in}
\affiliation{Department of Physics, Tezpur University, Tezpur-784028, India}
\author{Rinku Mishra}
\email{arra4723@gmail.com}
\affiliation{Department of Physics, Tezpur University, Tezpur-784028, India}

\begin{abstract}
We present an analysis of normal and inverted hierarchical neutrino mass models within the framework of tri-bi-maximal (TBM) mixing. Considering the neutrinos to be quasi-degenerate (QDN), we study two different neutrino mass models with mass eigenvalues $(m_1, -m_2, m_3)$ and $(m_1, m_2, m_3)$  for both normal hierarchical (NH) and inverted hierarchical (IH) cases. Parameterizing the neutrino mass matrix using best fit oscillation and cosmology data for a QDN scenario, we find the right-handed Majorana mass matrix using type I seesaw formula for two types of Dirac neutrino mass matrices: charged lepton (CL) type and up quark (UQ) type. Incorporating the presence of type II seesaw term which arises naturally in generic left-right symmetric models (LRSM) along with type I term, we compare the predictions for neutrino mass parameters with the experimental values. Within such a framework and incorporating both oscillation as well as cosmology data, we show that QDN scenario of neutrino masses can still survive in nature with some minor exceptions. A viable extension of the standard model with an abelian gauged flavor symmetry is briefly discussed which can give rise to the desired structure of the Dirac and Majorana mass matrices.
\end{abstract}
\maketitle
\section{Introduction}
Recent neutrino oscillation experiments have provided significant amount of evidence which confirms the existence of the non-zero yet tiny neutrino masses \cite{PDG}. We know that the smallness of three Standard Model
neutrino masses \cite{PDG} can be naturally explained 
via seesaw mechanism. In general, such seesaw mechanism can be of three types : type I \cite{ti}, type II \cite{tii} and type III \cite{tiii}. All these mechanisms involve the inclusion of additional fermionic or scalar fields to generate tiny neutrino masses at tree level. Although these seesaw models can naturally explain the smallness of neutrino mass compared to the electroweak scale, we are still far away from understanding the origin of neutrino mass hierarchies as suggested by experiments. Recent neutrino oscillation experiments T2K \cite{T2K}, Double ChooZ \cite{chooz}, Daya-Bay \cite{daya} and RENO \cite{reno} provide the values of various neutrino oscillation parameters as follows:
$$ \Delta m_{21}^2=(7.12-8.20) \times 10^{-5} $$
$$ \lvert \Delta m_{31}^2 \rvert=(2.21-2.64)\times 10^{-3} $$
$$ \text{sin}^2\theta_{12}=0.27-0.37 $$
$$ \text{sin}^2\theta_{23}=0.37-0.67 $$ 
\begin{equation}
\text{sin}^2\theta_{13}=0.017-0.033.
\end{equation}
The above recent data have positive evidence for non-zero $\theta_{13}$ as well, which was earlier thought to be zero or negligibly small. The values of these mixing angles have non-trivial impact on the neutrino mass hierarchy as studied in a recent paper \cite{SV} where the author showed that the atmospheric angle $\theta_{23}$ is found to discriminate the possible hierarchies in the type I and type II seesaw frameworks using different texture zero mass matrices. In this context, to know the actual hierarchy of the neutrino masses has become equally important like the issue of non-zero $\theta_{13}$ both from neutrino physics as well as cosmology point of view. Recent cosmological upper bound \cite{SFO} on the sum of three absolute neutrino masses $\sum_i m_i \leq 0.28\; \text{eV}$ has  ruled out Quasi-Degenerate Neutrino (QDN) mass models with $m_i \geq  0.1 \; \text{eV}$. This has made studying the survivability of QDN models as important as the issue of normal and inverted hierarchical nature of neutrino masses.

Detailed analysis of normal versus inverted hierarchical neutrino masses using different approaches started just after the discovery of the neutrino oscillation phenomena. Inverted hierarchical neutrino was studied exclusively in \cite{SP} considering neutrino as a pseudo Dirac particle with non-conservation of $L_e-L_{\mu}-L_{\nu}$ where $L_l$ denotes lepton number corresponding to individual lepton $(l)$ generation. Use of specific grand unified models explaining the seesaw mechanisms has also been done in the last few years to study the hierarchy of neutrino masses. An analysis done in \cite{CHA} showed that every normal neutrino mass hierarchy solution of a grand unified model corresponds to an inverted hierarchy solution. It was also mentioned in their work that any future observation of inverted hierarchy would tend to disfavor the grand unified models based on the conventional type I seesaw mechanism. But models with type II and type III or models based on conserved $ L_e-L_{\mu}-L_{\nu}$ symmetry may favor the inverted hierarchical nature of neutrino masses. Models based on seesaw mechanism with three right handed neutrinos can also generate inverted hierarchical neutrino masses \cite{NNS} within the framework of bi-maximal mixing. As stressed earlier, along with the hierarchy of neutrino masses, the explanation of non-zero $\theta_{13}$ as well as CP violation is also an unsolved agenda in neutrino physics. From supernova neutrinos point of view, it was shown \cite{HM} that one can discriminate the inverted hierarchy from the normal one if $\text{sin}^2\theta_{13}\geq \text{a few} \times 10^{-4}$. If a particular neutrino mass hierarchy is assumed this can bias cosmological parameter constraints \cite{PRD80} like dark energy equation of state parameter as well as the sum of the neutrino masses. 

In view of the importance of understanding the hierarchies as stressed above, this work is planned to present an analysis of normal and inverted neutrino mass hierarchies incorporating the contributions from both type I and type II seesaw mechanisms. However, our analysis do not attempt to explain non-zero $\theta _{13}$ as has already been explained in the literature by considering deviations from the TBM mixing using different corrections. The present analysis is done within the framework of TBM mixing which theoretically, is a very close approximate description of neutrino mixing. In the present analysis we use the appropriate neutrino mass patterns in the framework of TBM mixing by considering neutrinos to be QDN. The mass matrices are parameterized for QDN case with the help of present neutrino oscillation data and the cosmological upper bound on the sum of neutrino masses. Using these QDN mass matrices and considering two possible structures of the Dirac neutrino mass matrix $m_{LR}$: charged lepton (CL) type and up quark (UQ) type, we calculate the right-handed Majorana mass matrix using type I seesaw formula. We then take into account the contributions from type II seesaw term in a generic left-right symmetric theory \cite{lrsm}. In the presence of both type I and type II seesaw contributions, we perform our detailed analysis to calculate the predictions for neutrino parameters to show the survivability of the QDN scenario. In the end, we also outline a simple extension of standard model by an abelian gauged flavor symmetry which can give rise to the specific structure of Dirac neutrino mass matrices used in the analysis.

This paper is organized as follows: in section \ref{method} we discuss the methodology of type II seesaw mechanism. In section \ref{numeric} we discuss our numerical analysis and results. In section \ref{abelian} we outline a simple extension of standard model by an abelian gauged flavor symmetry which can naturally give rise to the desired structure of mass matrices and then finally conclude in section \ref{conclude}.
\section{Methodology and Type II seesaw mechanism }
\label{method}
Type I seesaw framework is the  simplest mechanism for generating tiny neutrino masses and mixing. There is also another type of non-canonical seesaw formula (known as type-II seesaw formula)\cite{tii} where  a left-handed Higgs triplet $\Delta_{L}$ picks up a vacuum expectation value (vev). This is possible both in the minimal extension of the standard model by $\Delta_{L}$ or in other well motivated extensions like left-right symmetric model (LRSM) \cite{lrsm}. The seesaw formula can be written as
\begin{equation}
m_{LL}=m_{LL}^{II} + m_{LL}^I
\label{type2a}
\end{equation}
where the usual type I seesaw formula  is given by the expression,
\begin{equation}
m_{LL}^I=-m_{LR}M_{RR}^{-1}m_{LR}^{T}.
\end{equation}
Here  $m_{LR}$ is the Dirac neutrino mass matrix. The above seesaw formula with both type I and type II contributions can naturally arise in extension of standard model with three right handed neutrinos and one copy of $\Delta_{L}$. However, we will use this formula in the framework of LRSM where $M_{RR}$ arises naturally as a result of parity breaking at high energy and both the type I and type II terms can be written in terms of $M_{RR}$ as we will see below.

In this present analysis, we consider $m_{LR}$ in a diagonal form and $M_{RR}$ in general non-diagonal form. In LRSM with Higgs triplets, $M_{RR}$ can be expressed as $M_{RR}=v_{R}f_{R}$ with $v_{R}$ being the vev of the right handed triplet Higgs field $\Delta_R$ imparting Majorana masses to the right-handed neutrinos and $f_{R}$ is the corresponding Yukawa coupling.
The first term $m_{LL}^{II}$ in equation (\ref{type2a}) is due to the vev of $SU(2)_{L}$ Higgs triplet. In the usual LRSM, $m_{LL}^{II}$ and $M_{RR}$ are proportional to the vev's of the electrically neutral components of scalar Higgs triplets $\Delta_L$ and $\Delta_R$ respectively. Thus, $m_{LL}^{II}=f_{L}v_{L}$ and $M_{RR}=f_{R}v_{R}$, where $v_{L,R}$ denote the vev's and $f_{L,R}$ are symmetric $3\times3$ matrices. The left-right symmetry demands $f_{R}=f_{L}=f$. The induced vev for the left-handed triplet $v_{L}$ can be shown for generic LRSM to be
$$v_{L}=\gamma \frac{M^{2}_{W}}{v_{R}}$$
with $M_{W}\simeq 80.4$ GeV being the weak boson mass such that 
$$ |v_{L}|<<M_{W}<<|v_{R}| $$ 
In general $\gamma$ is a function of various couplings in the scalar potential of generic LRSM and without any fine tuning $\gamma$ is expected to be of the order unity ($\gamma\sim 1$). Type II seesaw formula in equation (\ref{type2a}) can now be expressed as
\begin{equation}
m_{LL}=\gamma (M_{W}/v_{R})^{2}M_{RR}-m_{LR}M^{-1}_{RR}m^{T}_{LR}
\label{type2}
\end{equation}

With above seesaw formula (\ref{type2}), the neutrino mass matrices
are constructed by considering contributions from both type I and type II terms. Here, $M_{RR}$ is defined as $M_{RR}=v_{R}f_R$ . If $f_R$ is held
fixed, both terms in equation (\ref{type2}) vary as $1/v_R$. Here we hold $M_{RR}$ fixed, so the first term is $v_R$ dependent while second term is fixed. However, different choices of
$v_{R}$ for fixed $M_{RR}$ would lead to different values of $m_{LL}^{II}$ while keeping $m_{LL}^I$ unchanged. This ambiguity is seen in the literature where different choices of $v_{R}$ are made according to convenience \cite{bstn,bd,ca,aw}. However, in this present work we will always take $v_{R}$ as $v_R=\gamma\frac{M^2_W}{v_L} \simeq \gamma \times 10^{15}\;\text{GeV}$ \cite{aw}. It is worth mentioning that, here $SU(2)_R \times U(1)_{B-L}$ gauge symmetry breaking scale (as in generic LRSM) $v_R$ is the same as the scale of parity breaking \cite{bstn}. Using this form of $v_R$, the seesaw formula (\ref{type2}) becomes 
\begin{equation}
m_{LL}=\gamma \left (\frac{M_{W}}{\gamma \times 10^{15}} \right )^{2}M_{RR}-m_{LR}M^{-1}_{RR}m^{T}_{LR}
\label{type2b}
\end{equation}
Since type II term is inversely proportional to $\gamma$, smaller values of this parameter (say, $\gamma \sim 0$) would give rise to more dominating type II term whereas $\gamma \sim 1$ would correspond to the minimum possible contributions from type II term.

After fixing the symmetry breaking scales as above, we carry out a complete analysis of the normal and inverted hierarchical models of neutrino masses in the framework of TBM mixing. We vary the dimensionless parameter $\gamma$ from $0.001$ to $1.0$  and check the survivability of neutrino mass models with contributions from type I and type II terms. We adopt a \textit{natural selection} for the survival of neutrino mass models which have the least deviation of $\gamma$ from unity. Nearer the value of $\gamma$ to one, better the chance for the survival of the model in question. Thus the value of  $\gamma$ is an important parameter for the proposed natural selection of the neutrino mass models in question.

\section{Numerical analysis and results}
\label{numeric}

For detail numerical analysis we use the specific $\mu-\tau$ symmetric neutrino mass matrix \cite{ga} which gives rise to TBM type mixing pattern

\begin{equation}
m_{LL}=\left(\begin{array}{ccc}
A& B&B\\
B&\frac{1}{2}(A+B+D) &\frac{1}{2}(A+B-D) \\
B& \frac{1}{2}(A+B-D)& \frac{1}{2}(A+B+D)
\end{array}\right)
\label{matrix1}
\end{equation}
which has eigenvalues $m_1=A-B$, $m_2=A+2B$ and $m_3=D$.
Then we parameterize the above matrix for QDN case. From presently available cosmological constraints, the upper bound on sum of neutrino masses has come down to the lowest value $\sum_im_i\le0.28\;\text{eV}$ \cite{SFO} which has ruled out QDN neutrino models with $m_i\ge 0.1 \;\text{eV}$. Parametrization  of the matrix (\ref{matrix1}) is done with this upper bound and taking the largest allowed value $m_i\le 0.1 \;\text{eV}$ consistent with the latest cosmological data. A classification for three-fold QDN neutrino masses \cite{a} with maximum Majorana CP violating phase in their eigenvalues is used here. CP phase patterns in the mass eigenvalues for both NH and IH are taken as: $(m_1, -m_2, m_3)$ (denoted as $+-+$) and $(m_1, m_2, m_3)$ (denoted as $+++$). Using the best global fit values of neutrino oscillation observational data \cite{od} on solar and atmospheric neutrino mass squared differences, and taking $m_i\le 0.1 \; \text{eV}$, predictions for neutrino parameters are calculated within the cosmological upper bounds mentioned above. First, we calculate the neutrino mass parameters using the above form of the matrix (\ref{matrix1}) taking into account only type I seesaw contributions. These predictions along with the input parameters for IH and NH cases in presence of only type I seesaw are presented in Table \ref{table:results1}. Then we take into account contributions from type II seesaw term given in equation (\ref{type2b}) to study the survivability of the neutrino mass models. 
\begin{table}[ht]
\centering
\caption{Input parameters and Predictions for different parameters consistent with experiments using type I seesaw only}
\vspace{0.5cm}
\begin{tabular}{|c|c|c|c|c|}
 \hline
   Parameters & IH(+-+) &  IH(+++) &  NH(+-+)&  NH(+++)\\ \hline
$\Delta m_{21}^2[10^{-5}\; \text{eV}^2]$&7.65&7.65&7.65&7.65\\  \hline
$\lvert \Delta m_{13}^2\rvert[10^{-3} \; \text{eV}^2]$&2.40&2.40&2.40&2.40\\  \hline
$m_3\;(\text{eV})$&0.08&0.08&0.10&0.10\\  \hline
$\text{sin}^2\theta_{23}$&0.50&0.50&0.50&0.50\\  \hline
$\text{sin}^2\theta_{12}$&0.33&0.33&0.33&0.33\\  \hline
$m_1\;(\text{eV})$&0.09340&0.09340&0.08674&0.08675\\  \hline
$m_2\; (\text{eV})$&-0.09380&0.09380&-0.08717&0.08717\\  \hline
$\sum_i m_i\;(\text{eV})$&0.267&0.267&0.274&0.274\\  \hline
$A$ & 0.031&0.09353&0.02877&0.08688 \\ \hline
$B$&  -0.0624&0.00013&-0.05797&0.00014 \\  \hline 
$D$ &0.08&0.08&0.10&0.10\\ \hline
\end{tabular}
\label{table:results1}
\end{table}
 \begin{table}[ht]
\centering
\caption{Right Handed Majorana Neutrino masses in $\text{GeV}$}
\vspace{0.5cm}
\begin{tabular}{|c|c|c|}
 \hline

   $(m,n)$ &$ (6,2)$ & $(8,4)$ \\ \hline
   IH(+++) &  $\left(\begin{array}{ccc}
3.8\times 10^5 & -2.26\times 10^5 &-4.67\times 10^6\\
  -2.26\times 10^5 &7.23\times 10^{10} &-1.18\times 10^{11} \\
-4.67\times 10^6&-1.18\times 10^{11} &3.09\times 10^{13}
\end{array}\right)$& $\left(\begin{array}{ccc}

-2.19\times 10^3 &-1.34\times 10^3 &-5.75\times 10^{5}  \\
-1.34\times 10^3&4.32\times 10^8 &-1.45\times 10^{10} \\
-5.75\times 10^{5} &-1.45\times 10^{10}  &7.87\times 10^{13}
\end{array}\right)$  \\ \hline
    IH(+-+) &  $\left(\begin{array}{ccc}
123061 & -1.043\times 10^8 &-2.16\times 10^9\\
 -1.04\times 10^8 &2.80\times 10^{10} &-1.04\times 10^{12} \\
-2.16\times 10^9&-1.04\times 10^{12} &1.20\times 10^{13}
\end{array}\right)$& $\left(\begin{array}{ccc}

733.40& -6.22\times 10^5  & -2.66\times 10^8 \\
-6.22\times 10^5 &1.67\times 10^8 &-1.28\times 10^{11} \\
 -2.66\times 10^8&-1.28\times 10^{11}  &3.04\times 10^{13}
\end{array}\right)$ \\  \hline 
NH(+++) & $\left(
\begin{array}{ccc}
 3.95\times 10^5 & -2.91\times 10^5. & -6.01\times 10^6 \\
 -2.91\times 10^5 & 6.72\times 10^{10} & 9.64\times 10^{10} \\
 -6.01\times 10^6 & 9.64\times 10^{10} & 2.87\times 10^{13}
\end{array}
\right)$& $\left(
\begin{array}{ccc}
 2.36\times 10^3 & -1.73\times 10^3& -7.40\times 10^5 \\
 -1.73\times 10^3& 4.01\times 10^8 & 1.19\times 10^{10} \\
 -7.40\times 10^5 & 1.19\times 10^{10} & 7.3\times 10^{13}
\end{array}
\right)$\\ \hline
NH(+-+)& $\left(\begin{array}{ccc}
 1.3\times 10^5 & -1.12\times 10^8 & -2.32\times 10^9 \\
 -1.12\times 10^8 & 1.93\times 10^{10} & -8.92\times 10^{11} \\
 -2.32\times 10^9 & -8.92\times 10^{11} & 8.27\times 10^{12}
\end{array}
\right)$& $\left(\begin{array}{ccc}
 790.241 & -6.70\times 10^5 & -2.86\times 10^8 \\
 -6.70\times 10^5 & 1.16\times 10^8 & -1.10\times 10^{11} \\
 -2.86\times 10^8 & -1.10\times 10^{11} & 2.10\times 10^{13}
\end{array}\right)$\\  \hline

\end{tabular}
\label{table:results2}
\end{table}
Using the inverse type I seesaw formula 
\begin{equation}
M_{RR}=m_{LR}^Tm_{LL}^{-1}m_{LR} 
\end{equation}
first we calculate the $M_{RR}$ for each case using Dirac neutrino mass $(m_{LR})$ in the diagonal form. In this analysis $m_{LR}$ is being taken as either the charged lepton mass matrix or up quark mass matrix. The general form of  Dirac neutrino mass is  
\begin{equation}
m_{LR}=\left(\begin{array}{ccc}
\lambda^m & 0 & 0\\
0 & \lambda^n & 0 \\
0 & 0 & 1
\end{array}\right)m_f
\label{mLR1}
\end{equation}
where $m_f$ corresponds to $m_\tau \tan{\beta}$ for $(m, n) = (6, 2), \; \tan{\beta} = 40$ in case of charged lepton and $m_t$ for $(m, n) = (8, 4)$ in the case of up quarks  \cite{dm,mkd}. $\lambda = 0.22$ is the standard Wolfenstein parameter. $M_{RR}$ for both the cases are presented in the Table \ref{table:results2}. 

\begin{figure}[ht]
 \centering
\includegraphics{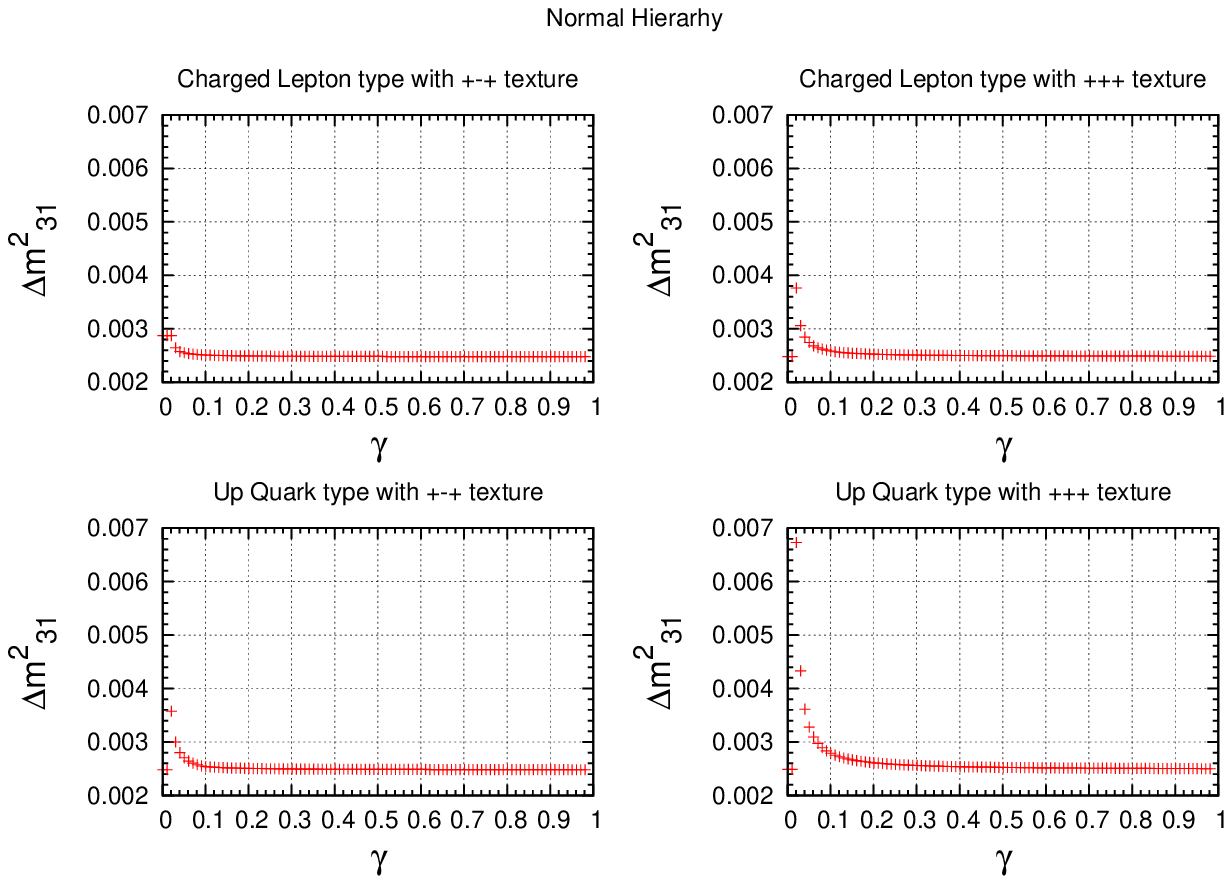}
\caption{Variation of the predicted values of $\Delta m^2_{31}$ as a function of $\gamma$ in NH case for both charged lepton and up quark type $m_{LR}$ as well as both types of maximal Majorana CP phases}
\label{fig1}
\end{figure}

\begin{figure}[ht]
 \centering
\includegraphics{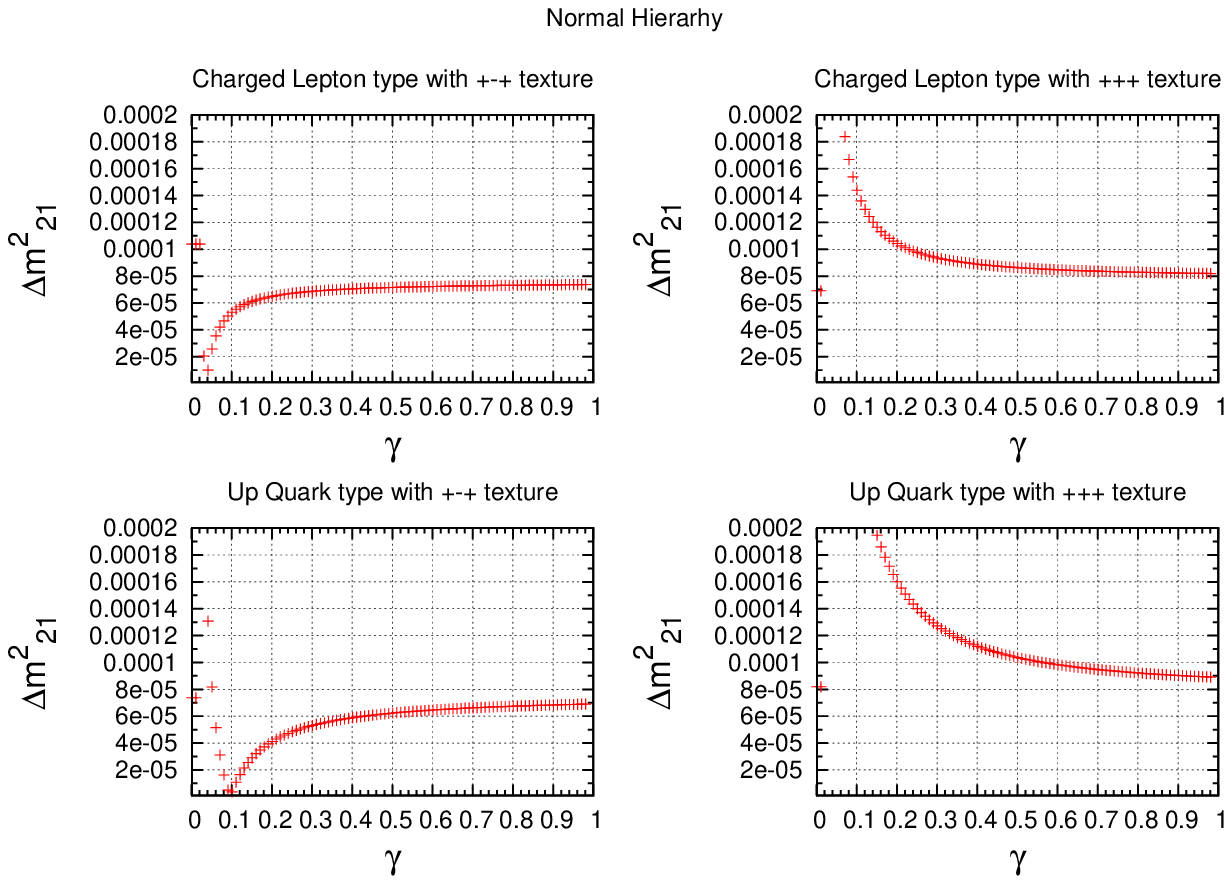}
\caption{Variation of the predicted values of $\Delta m^2_{21}$ as a function of $\gamma$ in NH case for both charged lepton and up quark type $m_{LR}$ as well as both types of maximal Majorana CP phases}
\label{fig2}
\end{figure}

\begin{figure}[ht]
 \centering
\includegraphics{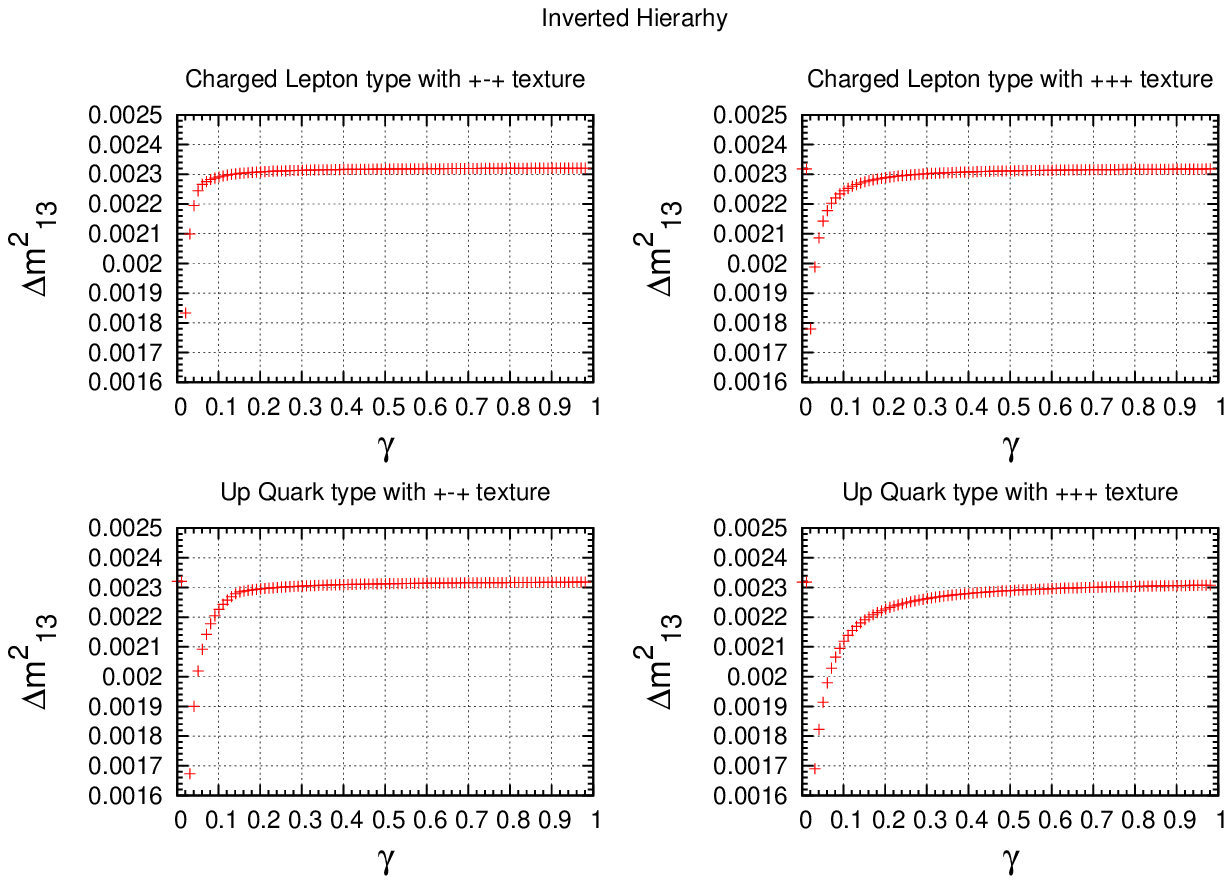}
\caption{Variation of the predicted values of $\Delta m^2_{13}$ as a function of $\gamma$ in IH case for both charged lepton and up quark type $m_{LR}$ as well as both types of maximal Majorana CP phases}
\label{fig3}
\end{figure}

\begin{figure}[ht]
 \centering
\includegraphics{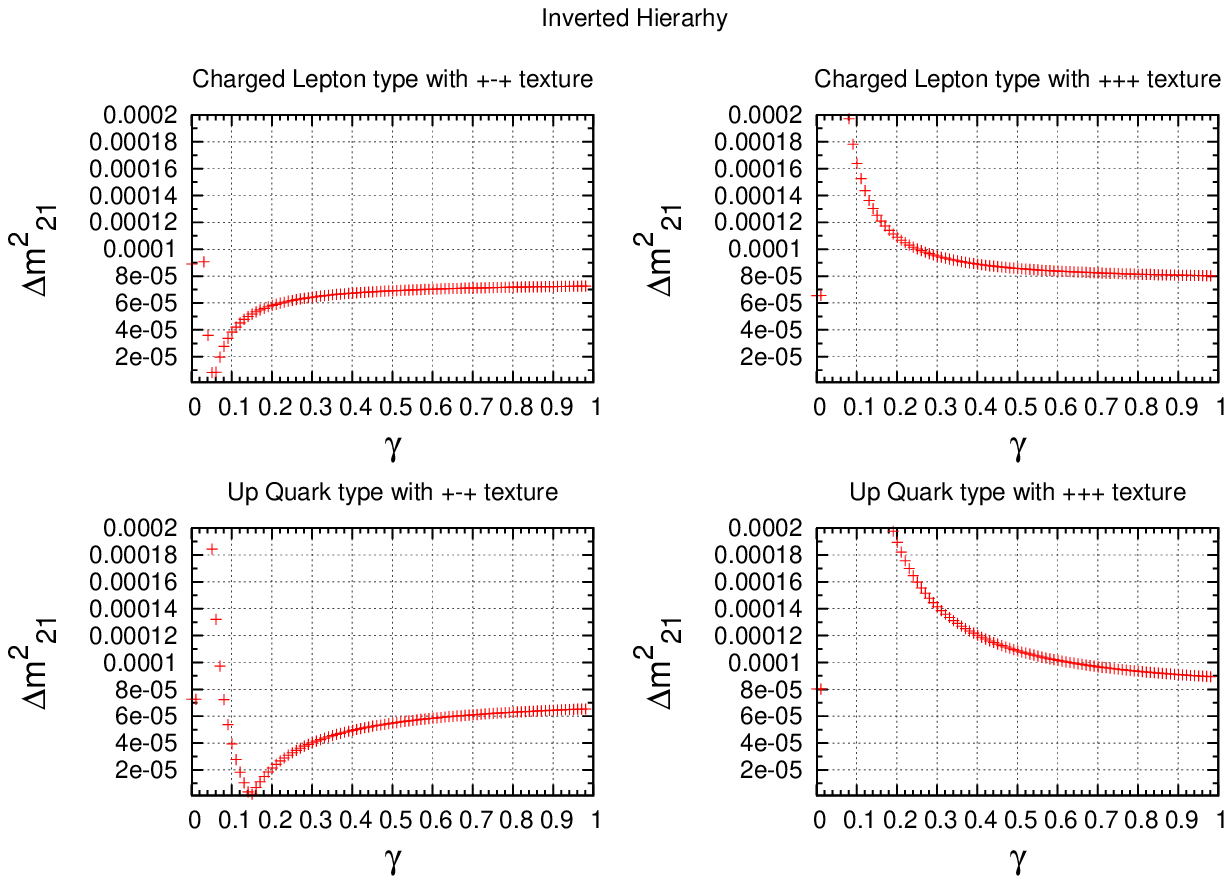}
\caption{Variation of the predicted values of $\Delta m^2_{21}$ as a function of $\gamma$ in IH case for both charged lepton and up quark type $m_{LR}$ as well as both types of maximal Majorana CP phases}
\label{fig4}
\end{figure}

Entering the values of $M_{RR}$ in the equation (\ref{type2b}), we compute the type I+II neutrino mass matrix $m_{LL}$ and find the mass eigenvalues and eigenvectors to compute the $\lvert\Delta m^2_{31}\rvert$ and $\Delta m^2_{21}$ and corresponding mixing angles for various values of $\gamma$. Deviations of these predicted $\Delta m^2$'s from the data central values are then plotted in Figs. \ref{fig1}, \ref{fig2}, \ref{fig3}, \ref{fig4} against the parameter $\gamma$. It is observed from the figures that all the mass models survive at $\gamma \sim 1$ except the fact that for UQ type $m_{LR}$, the predictions for $\Delta m_{21}^2$ has deviated slightly from the $3\sigma$ range of experimental data. On the other hand, predictions for  the mixing angles i.e. $\theta_{12}$, $\theta_{23}$ show that all the models survive provided $\gamma$ is close to $1$ (or in other words, type II term has minimal contribution). The calculated values of neutrino parameters for  $\gamma =0.25, 0.50, 0.75, 1.0$ are given in Table \ref{table:results3}, \ref{table:results4}, \ref{table:results5} and \ref{table:results6}. From this analysis we observe that QDN neutrino mass models can survive in nature within the framework of TBM mixing (with a few exceptions), with contributions from both type I and type II seesaw mechanisms.

\begin{table}[ht]
\centering
\caption{Predictions for neutrino parameters using type I+II seesaw for CL type $m_{LR}$ with Inverted Hierarchy}
\vspace{0.5cm}
\begin{tabular}{|c|c|c|c|c|c|c|c|c|}
 \hline
   Parameters & IH(+-+) & IH(+-+) & IH(+-+) & IH(+-+) & IH(+++) & IH(+++) & IH(+++) & IH(+++) \\
              & $\gamma = 0.25$ & $\gamma = 0.50$ & $\gamma = 0.75$ & $\gamma = 1.00$ & $\gamma = 0.25$ & $\gamma = 0.50$ & $\gamma = 0.75$ & $\gamma = 1.00$ \\ \hline
$\Delta m^2_{21}[10^{-5}\;\text{eV}^2]$ & 6.19 & 6.91 & 7.14 & 7.25 & 10.01 & 8.57 & 8.18 & 8.02\\ \hline
$ \lvert \Delta m^2_{31} \rvert [10^{-3}\;\text{eV}^2] $ & 2.31 & 2.32 & 2.32 & 2.32 & 2.29 & 2.31 & 2.31 & 2.31  \\ \hline   
$\text{sin}^2\theta_{23}$&0.50&0.50&0.50&0.50 & 0.51 & 0.50 & 0.50 & 0.50 \\  \hline
$\text{sin}^2\theta_{12}$&0.33&0.33&0.33&0.33 & 0.15 & 0.22 & 0.26 & 0.33 \\  \hline
\end{tabular}
\label{table:results3}
\end{table}
\begin{table}[ht]
\centering
\caption{Predictions for neutrino parameters using type I+II seesaw for CL type $m_{LR}$ with Normal Hierarchy}
\vspace{0.5cm}
\begin{tabular}{|c|c|c|c|c|c|c|c|c|}
 \hline
   Parameters & NH(+-+) & NH(+-+) & NH(+-+) & NH(+-+) & NH(+++) & NH(+++) & NH(+++) & NH(+++) \\
              & $\gamma = 0.25$ & $\gamma = 0.50$ & $\gamma = 0.75$ & $\gamma = 1.00$ & $\gamma = 0.25$ & $\gamma = 0.50$ & $\gamma = 0.75$ & $\gamma = 1.00$ \\ \hline
$\Delta m^2_{21}[10^{-5}\;\text{eV}^2]$ & 6.72 & 7.16 & 7.30 & 7.36 & 9.76 & 8.63 & 8.32 & 8.19\\ \hline
$ \lvert \Delta m^2_{31} \rvert [10^{-3}\;\text{eV}^2] $ & 2.48 & 2.48 & 2.48 & 2.47 & 2.51 & 2.49 & 2.49 & 2.48  \\ \hline   
$\text{sin}^2\theta_{23}$&0.49&0.49&0.50&0.50 & 0.48 & 0.49 & 0.49 & 0.50 \\  \hline
$\text{sin}^2\theta_{12}$&0.33&0.33&0.33&0.33 & 0.20 & 0.25 & 0.27 & 0.33 \\  \hline
\end{tabular}
\label{table:results4}
\end{table}
\begin{table}[ht]
\centering
\caption{Predictions for neutrino parameters using type I+II seesaw for UQ type $m_{LR}$ with Inverted Hierarchy}
\vspace{0.5cm}
\begin{tabular}{|c|c|c|c|c|c|c|c|c|}
 \hline
   Parameters & IH(+-+) & IH(+-+) & IH(+-+) & IH(+-+) & IH(+++) & IH(+++) & IH(+++) & IH(+++) \\
              & $\gamma = 0.25$ & $\gamma = 0.50$ & $\gamma = 0.75$ & $\gamma = 1.00$ & $\gamma = 0.25$ & $\gamma = 0.50$ & $\gamma = 0.75$ & $\gamma = 1.00$ \\ \hline
$\Delta m^2_{21}[10^{-5}\;\text{eV}^2]$ & 3.28 & 5.49 & 6.20 & 6.53 & 15.98 & 10.86 & 9.49 & 8.94\\ \hline
$ \lvert \Delta m^2_{31} \rvert [10^{-3}\;\text{eV}^2] $ & 2.30 & 2.31 & 2.31 & 2.31 & 2.24 & 2.28 & 2.30 & 2.30  \\ \hline   
$\text{sin}^2\theta_{23}$&0.50&0.50&0.50&0.50 & 0.54 & 0.52 & 0.51 & 0.50 \\  \hline
$\text{sin}^2\theta_{12}$&0.33&0.33&0.33&0.33 & 0.12 & 0.22 & 0.27 & 0.33 \\  \hline
\end{tabular}
\label{table:results5}
\end{table}
\begin{table}[ht]
\centering
\caption{Predictions for neutrino parameters using type I+II seesaw for UQ type $m_{LR}$ with Normal Hierarchy}
\vspace{0.5cm}
\begin{tabular}{|c|c|c|c|c|c|c|c|c|}
 \hline
   Parameters & NH(+-+) & NH(+-+) & NH(+-+) & NH(+-+) & NH(+++) & NH(+++) & NH(+++) & NH(+++) \\
              & $\gamma = 0.25$ & $\gamma = 0.50$ & $\gamma = 0.75$ & $\gamma = 1.00$ & $\gamma = 0.25$ & $\gamma = 0.50$ & $\gamma = 0.75$ & $\gamma = 1.00$ \\ \hline
$\Delta m^2_{21}[10^{-5}\;\text{eV}^2]$ & 4.83 & 6.23 & 6.68 & 6.89 & 14.00 & 10.36 & 9.33 & 8.90\\ \hline
$ \lvert \Delta m^2_{31} \rvert [10^{-3}\;\text{eV}^2] $ & 2.49 & 2.48 & 2.48 & 2.48 & 2.58 & 2.52 & 2.50 & 2.50  \\ \hline   
$\text{sin}^2\theta_{23}$&0.49&0.50&0.50&0.50 & 0.46 & 0.48 & 0.49 & 0.50 \\  \hline
$\text{sin}^2\theta_{12}$&0.33&0.33&0.33&0.33 & 0.15 & 0.20 & 0.26 & 0.33 \\  \hline
\end{tabular}
\label{table:results6}
\end{table}
\section{A viable model with a gauged abelian flavor symmetry}
\label{abelian}
The standard model (SM) of particle physics, if extended by the inclusion of three right handed neutrinos which are singlet under the SM gauge group, can give rise to tiny neutrino mass by type I seesaw mechanism \cite{ti}. Alternatively, if the SM is extended by a scalar triplet, tiny neutrino mass can arise from type II seesaw mechanism \cite{tii} after the neutral component of the scalar triplet acquires a tiny vacuum expectation value. Being singlets under the gauge group, the mass matrix of the right handed neutrinos can have off-diagonal terms as well. Similarly, the gauge structure of the SM does not prevent off-diagonal Dirac Yukawa couplings. In other words both the Dirac and right handed Majorana mass matrices can be non-diagonal in general. However, throughout our analysis in the previous sections, we have restricted the Dirac neutrino mass matrix to its diagonal form only. This can be achieved by incorporating additional symmetries (global or local) with family non-universal gauge charges so that off diagonal mass terms are not allowed. In this work, we take up one such highly motivated extended symmetry: abelian gauge extension of the SM. It should be noted that although our analysis in the previous sections considered the type II seesaw formula (\ref{type2}) for a general left-right symmetric model, here we outline a simpler extension of the standard model by an abelian gauged flavor symmetry to explain the desired form of the mass matrices.

Abelian gauge extension of Standard Model is one of the best motivating examples of beyond Standard Model physics. 
For a review see \cite{Langacker}. Such a model is also motivated within the framework of GUT models, for example $E_6$. 
The supersymmetric version of such models have an additional advantage in the sense that they provide a solution to the MSSM (Minimal Supersymmetric Standard Model) $\mu$ problem. Such abelian gauge extension of SM was studied recently in \cite{Borah} in the context of neutrino mass and cosmology.

Here we consider an extension of the Standard Model gauge group with one abelian $U(1)_X$ gauge symmetry. Thus, the model we propose here is an $SU(3)_c \times SU(2)_L \times U(1)_Y \times U(1)_X$ gauge theory with three chiral generations of SM and three additional right handed neutrinos. We will consider family 
non universal $U(1)_X$ couplings.

The fermion content of our model is 
\begin{equation}
Q_i=
\left(\begin{array}{c}
\ u \\
\ d
\end{array}\right)
\sim (3,2,\frac{1}{6},n_{qi}),\hspace*{0.8cm}
L_i=
\left(\begin{array}{c}
\ \nu \\
\ e
\end{array}\right)
\sim (1,2,-\frac{1}{2},n_{li}), \nonumber 
\end{equation}
\begin{equation}
u^c_i \sim (3^*,1,\frac{2}{3},n_{ui}), \quad d^c_i \sim (3^*,1,-\frac{1}{3},n_{di}), \quad e^c_i \sim (1,1,-1,n_{ei}), \quad \nu^c_i \sim (1,1,0,n_{ri}) \nonumber 
\end{equation}
where $ i=1,2,3 $ goes over the three generations of Standard Model and the numbers in the bracket correspond to the quantum number under the gauge group $SU(3)_c \times SU(2)_L \times U(1)_Y \times U(1)_X$. The $U(1)_X$ gauge quantum numbers should be such that they do not give rise to anomalies. We consider the following solution of the anomaly matching conditions
$$ n_{qi}=n_{ui}=n_{di}=0, \;\;\; n_{li}=n_{ei} = n_{ri} = n_i $$
$$ \sum n_{li} = \sum n_{ei} = \sum n_{ri} = 0, \;\;\;  \sum n^3_{li} = \sum n^3_{ei} = \sum n^3_{ri} = 0$$
In particular, if we choose $n_1 = 0, n_2 = n, n_3 = -n$, only the following types of Dirac Yukawa terms will be present in the Lagrangian.
$$ \mathcal{L}_Y \supset Y^{ii}_{\nu} \overline{L}_i H \nu^c_i+ Y^{ii}_e \overline{L}_i H^{\dagger} e^c_i $$
where $H$ is the Higgs field responsible for breaking electroweak symmetry and has the quantum numbers $(1,2,-\frac{1}{2},0)$ with respect to the gauge group. For the chosen abelian charges, two singlet Higgs fields must exist $S_1(1,1,0,0), S_2(1,1,0,2n)$ to give rise to a general structure of the right handed Neutrino mass matrix. One of these singlet fields $S_2(1,1,0,2n)$ (after acquiring a non-zero vev) also breaks the gauge symmetry $SU(3)_c \times SU(2)_L \times U(1)_Y \times U(1)_X$ to that of the standard model. Also, since the quarks have zero charges under the additional abelian symmetry, they continue to have usual CKM (Cabibbo Kobayashi Maskawa) structure of mixing matrix. Such a model can have rich phenomenology from collider as well as cosmology point of view. However, for the purpose of our current work, we outline this model just to explain one possible origin of the specific structure (diagonal) of Dirac neutrino mass matrix $m_{LR}$ used in the analysis. A more detailed investigation of such a model is left for future studies.

\section{Discussion}
\label{conclude}
Analysis of the effect of Majorana CP phases in case of quasi-degenerate neutrinos is done with contribution from both type I and type II seesaw formula within the framework of TBM mixing. Fitting the neutrino mass matrix with best fit oscillation and cosmology data, the right-handed neutrino mass matrix is calculated using type I seesaw formula only for both CL type and UQ type Dirac neutrino mass matrices. Adding type II seesaw term (which arises in generic left-right symmetric models) to the type I, the predictions for neutrino parameters are calculated. It is observed that for the minimal possible contribution of type II seesaw term (which corresponds to the value of the dimensionless parameter $\gamma$ in type II seesaw term of order 1) to the neutrino mass matrix, all the neutrino mass models can survive in nature except the fact that for UQ type $m_{LR}$ the predictions for $\Delta m_{21}^2$ has deviated slightly from the $3\sigma$ range of experimental data. Apart from this exception, all other predicted values of the neutrino parameters are consistent with neutrino oscillation data. Apart from neutrino oscillation data, these predictions are also within the limit of the cosmological upper bound $\sum_i m_i \le 0.28\; \text{eV}$. In view of above, the scenario of quasi-degenerate neutrinos can survive in nature within the framework of type I and type II seesaw mechanism and hence can not be ruled out yet. However, here we stick to the TBM mixing framework and have not made any attempt to explain the non-zero $\theta_{13}$ as confirmed recently by several neutrino oscillation experiments. As an extension of this work, one can incorporate various corrections to $m_{LL}$ to explain non-zero $\theta_{13}$ which we have left for future studies.

\end{document}